\documentclass[manuscript]{aastex}

\usepackage{graphics,color}
\usepackage{epstopdf}
\usepackage{ulem}
\definecolor{red}{rgb}{1,0,0}
\shorttitle{Super Fast Rotator: 2005 EC~EC$_{127}$}
\shortauthors{Chang et al.}
\begin{document}
\title{Confirmation of Large Super-Fast Rotator (144977) 2005~EC$_{127}$}

\author{Chan-Kao Chang\altaffilmark{1}; Hsing-Wen Lin\altaffilmark{1}; Wing-Huen Ip\altaffilmark{1,2}; Zhong-Yi Lin\altaffilmark{3}; Thomas Kupfer\altaffilmark{3}; Thomas A. Prince\altaffilmark{3}; Quan-Zhi Ye\altaffilmark{3,4}; Russ R. Laher\altaffilmark{4}; Hee-Jae Lee\altaffilmark{5,6}; Hong-Kyu Moon\altaffilmark{6}}

\altaffiltext{1}{Institute of Astronomy, National Central University, Jhongli,Taiwan}
\altaffiltext{2}{Space Science Institute, Macau University of Science and Technology, Macau}
\altaffiltext{3}{Division of Physics, Mathematics and Astronomy, California Institute of Technology, Pasadena, CA 91125, USA}
\altaffiltext{4}{Infrared Processing and Analysis Center, California Institute of Technology, Pasadena, CA 91125, U.S.A.}
\altaffiltext{5}{Department of Astronomy and Space Science, Chungbuk National University, 1, Chungdae-ro, Seowon-Gu, Cheongju, Chungbuk, 28644 Korea}
\altaffiltext{6}{Korea Astronomy and Space Science Institute, 776, Daedeok-daero, Yuseong gu, Daejeon, 305-348 South Korea}

\email{rex@astro.ncu.edu.tw}
\received{April 4 2017}
\accepted{April 27 2017}

\begin{abstract}
(144977) 2005~EC$_{127}$ is an V-/A-type inner-main-belt asteroid with a diameter of $0.6 \pm 0.1$~km. Asteroids of this size are believed to have rubble-pile structure, and, therefore, cannot have a rotation period shorter than 2.2 hours. However, our measurements show that asteroid 2005~EC$_{127}$ completes one rotation in $1.65 \pm 0.01$~hours with a peak-to-peak light-curve variation of $\sim 0.5$~mag. Therefore, this asteroid is identified as a large super-fast rotator. Either a rubble-pile asteroid with a bulk density of $\rho \sim 6$~g~cm$^{-3}$ or an asteroid with an internal cohesion of $47 \pm 30$~Pa can explain 2005~EC$_{127}$. However, the scenario of high bulk density is very unlikely for asteroids. To date, only six large super-fast rotators, including 2005~EC$_{127}$, have been reported, and this number is very small when compared with the much more numerous fast rotators. We also note that none of the six reported large SFRs are classified as C-type asteroids.
\end{abstract}


\section{Introduction}
The large (i.e., a diameter of a few hundreds of meters) super-fast rotators (hereafter, SFRs) are of interest for understanding asteroid interior structure. Because asteroids of sub-kilometer size are believed to have rubble-pile structure (i.e., gravitationally bounded aggregations) and cannot have super-fast rotation, defined as a rotation period shorter than 2.2 hours~\citep{Harris1996}\footnote{The 2.2-hour spin barrier was calculated for an asteroid with a bulk density of $\rho = 3$~g~cm$^{-3}$.}. However, the first large SFR, 2001~OE84, a near-Earth asteroid of $\sim 0.7$~km in size and completing one rotation in 29.19 minutes \citep{Pravec2002}, cannot be explained by rubble-pile structure, and, consequently, internal cohesion was proposed to be a possible solution \citep{Holsapple2007}. Although several attempts were made to discover large SFRs with extensive-sky surveys \citep{Masiero2009, Dermawan2011}, this asteroid group was not confirmed until another large SFR, 2005~UW163, was found by \citet{Chang2014b}. Up to now, five large SFRs have been reported, additionally including 1950~DA \citep{Rozitis2014}, 2000~GD65 \citep{Polishook2016}, and 1999~RE88 \citep{Chang2016}. However, the population size of large SFRs is still not clear. Compared with the 738 large fast rotators (i.e., diameters between 0.5--10 km and rotation periods between 2-3 hours) in the up-to-date Asteroid Light Curve Database \citep[hereafter, LCDB\footnote{http://www.minorplanet.info/lightcurvedatabase.html};][]{Warner2009}, large SFRs are rare. Either the difficulty of discovering them due to their sub-kilometer sizes (i.e., relatively faint) or the intrinsically small population size of this group could lead to this rarity in detection. Therefore, a more comprehensive survey of asteroid rotation period with a wider sky coverage and a deeper limiting magnitude, such as the ZTF\footnote{Zwicky Transient Facility; http://ptf.caltech.edu/ztf}, could help in finding more large SFRs. With more SFR samples, a thorough study of their physical properties could be conducted, and, therefore, further insights about asteroid interior structure are possible. To this objective, the TANGO project\footnote{Taiwan New Generation OIR Astronomy} has been conducting asteroid rotation-period surveys since 2013 using the iPTF\footnote{intermediate Palomar Transient Factory; http://ptf.caltech.edu/iptf} \citep[for details, see][]{Chang2014a, Chang2015, Chang2016}. From these surveys, two large SFRs and 27 candidates were discovered. Here we report the confirmation of asteroid (144977) 2005~EC$_{127}$ as a new large SFR. The super-fast rotation of (144977) 2005~EC$_{127}$ was initially and tentatively identified in the asteroid rotation-period survey using the iPTF in Feb 2015 \citep{Chang2016}, and then later confirmed in this work by follow-up observations using the Lulin One-meter Telescope in Taiwan \citep[LOT;][]{Kinoshita2005}.

This article is organized as follows. The observations and measurements are given in Section 2, the rotation period analysis is described in Section 3, the results and discussion are presented in Section 4, and a summary and conclusions can be found in Section 5.

\section{Observations}

The iPTF, LOT, and spectroscopic observations that support the findings in this work are described in this section.  The details of each of these observation runs are summarized in Table~\ref{obs_log}.

\subsection{iPTF Observations}\label{PTF_obs}
The iPTF is a follow-on project of the PTF, a project whose aim is to explore the transient and variable sky synoptically.  The iPTF/PTF employ the Palomar 48-inch Oschin Schmidt Telescope and an 11-chip mosaic CCD camera with a field of view of $\sim7.26$~deg$^2$ \citep{Law2009, Rau2009}. This wide field of view is extremely useful in collecting a large number of asteroid light curves within a short period of time. Four filters are currently available, including a Mould-{\it R}, Gunn-{\it g'}, and two different $H \alpha$ bands. The exposure time of the PTF is fixed at 60 seconds, which routinely reaches a limiting magnitude of $R \sim $21 mag at the $5\sigma$ level \citep{Law2010}. All iPTF exposures are processed by the IPAC-PTF photometric pipeline \citep{Grillmair2010, Laher2014}, and the Sloan Digital Sky Survey fields \citep[SDSS;][]{York2000} are used in the magnitude calibration. Typically, an accuracy of $\sim 0.02$~mag can be reached for photometric nights \citep{Ofek2012a, Ofek2012b}. Since the magnitude calibration is done on a per-night, per-filter, per-chip basis, small photometric zero-point variations are present in PTF catalogs for different nights, fields, filters and chips.

In the asteroid rotation-period survey conducted on Feb 25--26, 2015, we repeatedly observed six consecutive PTF fields near the ecliptic plane, in the $R$-band with a cadence of $\sim10$~minutes. Asteroid 2005~EC$_{127}$ was observed in the PTF field centered at $R.A. = 154.04^{\circ}$ and $Dec. = 10.12^{\circ}$ when it was approaching its opposition at a low phase angle of $\alpha \sim 1.3^{\circ}$. After all stationary sources were removed from the source catalogs, the light curves for known asteroids were extracted using a radius of 2\arcsec\ to match with the ephemerides obtained from the {\it JPL/HORIZONS\/} system. The light curve of 2005~EC$_{127}$ contains 42 clean detections from this observation run (i.e., the detections flagged as defective by the IPAC-PTF photometric pipeline were not included in the light curve).

\subsection{LOT Observations}
The follow-up observations to confirm the rotation period of 2005~EC$_{127}$ were carried out on Sept 24, 2016 using the LOT when 2005~EC$_{127}$ had a magnitude of $r' \sim 19.2$ at its low phase angle of $\alpha \sim 2.6^{\circ}$. The average seeing during the observations was $\sim1.3$\arcsec. All images were taken in the $r'$-band with a fixed exposure time of 300 seconds using the Apogee U42 camera, a 2K$\times$2K charge-coupled device with a pixel scale of 0.35\arcsec. We acquired a total of 84~exposures over a time span of $\sim 440$~minutes, and the time difference between consecutive exposures was $\sim 5$~minutes. The image processing and reduction included standard procedures of bias and flat-field corrections, astrometric calibration using $astrometry.net$\footnote{http://astrometry.net}, and aperture photometry using SExtractor \citep{Bertin1996}. The photometric calibration was done against Pan-STARRS1 point sources of $r' \sim 14$ to 22 mag \citep{Magnier2016} using linear least-squares fitting, which typically achieved a fitting residual $\sim 0.01$~mag. We improved the photometric accuracy by employing the trail-fitting method \citep{Veres2012, Lin2015} to accommodate the streaked image of 2005~EC$_{127}$ as a result of asteroid motion over the 300-second exposure time.

\subsection{Spectroscopic Observations}
\begin{figure}
\plotone{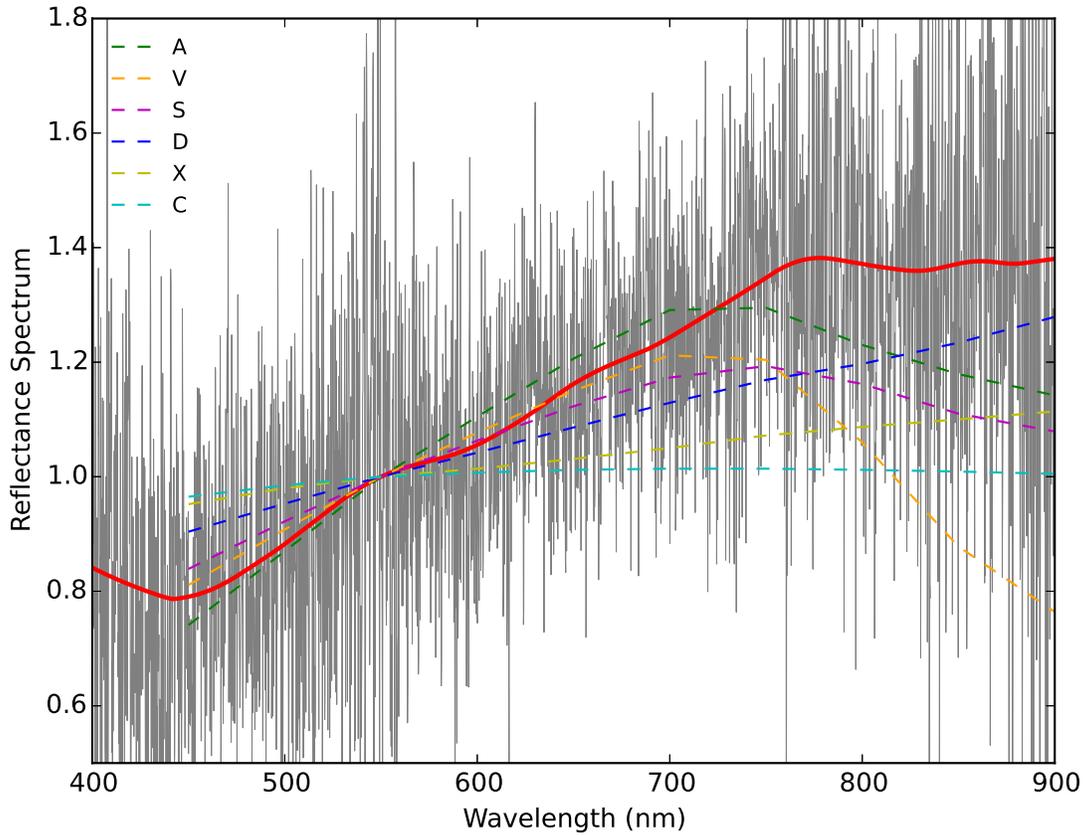}
\caption{Reflectance spectrum of 2005~EC$_{127}$ taken by the P200. The gray line is the original reflectance spectrum and the red line is the running average using locally weighted scatter-plot smoothing \citep[LOWESS;][]{Cleveland1979}. The colored dashed lines are the reference spectra of A- (green), V- (orange), S- (magenta), D- (blue), X- (yellow), and C-type (cyan) asteroids obtained from \citet{DeMeo2009}. All spectra are normalized at wavelength 500~nm.}
\label{spec}
\end{figure}

To determine the taxonomic type for 2005~EC$_{127}$, its optical spectra were obtained using the Palomar 200-inch Hale Telescope (hereafter, P200) and the Double-Beam Spectrograph \citep[DBSP;][]{Oke1982} in low-resolution mode ($R\sim1500$). Three consecutive exposures were taken on Oct 4, 2016 with an exposure time of 300 seconds each. An average bias frame was made out of 10 individual bias frames and a normalized flat-field frame was constructed out of 10 individual lamp flat-field exposures. For the blue and red arms, respectively, FeAr and HeNeAr arc exposures were taken at the beginning of the night. Both arms of the spectrograph were reduced using a custom \texttt{PyRAF}-based pipeline\footnote{https://github.com/ebellm/pyraf-dbsp} \citep{Bellm2016}. The pipeline performs standard image processing and spectral reduction procedures, including bias subtraction, flat-field correction, wavelength calibration, optimal spectral extraction, and flux calibration. The average spectrum of 2005~EC$_{127}$ was constructed by combining all individual exposures, and then it was divided by the solar spectrum\footnote{The solar spectrum was obtained from \citet{Kurucz1984}, and was then convolved with a Gaussian function to match the resolution of the spectrum of 2005~EC$_{127}$.} to obtain the reflectance spectrum of 2005~EC$_{127}$ (Fig.~\ref{spec}). The trend of the reflectance spectrum suggests an V-/A-type asteroid for 2005~EC$_{127}$, according to the Bus-DeMeo classification scheme~\citep{DeMeo2009}.

\section{Rotation-Period Analysis}
Before measuring the synodic rotation period for 2005~EC$_{127}$, the light-curve data points were corrected for light-travel time, and were reduced to both heliocentric ($r$) and geocentric ($\Delta$) distances at 1 AU by $M = m + 5\log(r\Delta)$, where $M$ and $m$ are reduced and apparent magnitudes, respectively. A second-order Fourier series \citep{Harris1989} was then applied to search for the rotation periods:
\begin{equation}\label{FTeq}
  M_{i,j} = \sum_{k=1,2}^{N_k} B_k\sin\left[\frac{2\pi k}{P} (t_j-t_0)\right] + C_k\cos\left[\frac{2\pi k}{P} (t_j-t_0)\right] + Z_i,
\end{equation}
where $M_{i,j}$ is the reduced magnitude measured at the light-travel-time-corrected epoch, $t_j$; $B_k$ and $C_k$ are the Fourier coefficients; $P$ is the rotation period; $t_0$ is an arbitrary epoch; and $Z_i$ is the zero point. For the PTF light curve, the fitting of $Z_i$ also includes a correction for the small photometric zero-point variations mentioned in Section~\ref{PTF_obs} \citep[for details, see][]{Polishook2012}. To obtain the other free parameters for a given $P$, we used least-squares minimization to solve Eq.~(\ref{FTeq}). The frequency range was explored between 0.25--50 rev/day with a step of 0.001 rev/day. To estimate the uncertainty of the derived rotation periods, we calculated the range of periods with $\chi^2$ smaller than $\chi_{best}^2+\triangle\chi^2$, where $\chi_{best}^2$ is the chi-squared value of the picked-out period and $\triangle\chi^2$ is obtained from the inverse chi-squared distribution, assuming $1 + 2N_k + N_i$ degrees of freedom.

\begin{figure}
\plotone{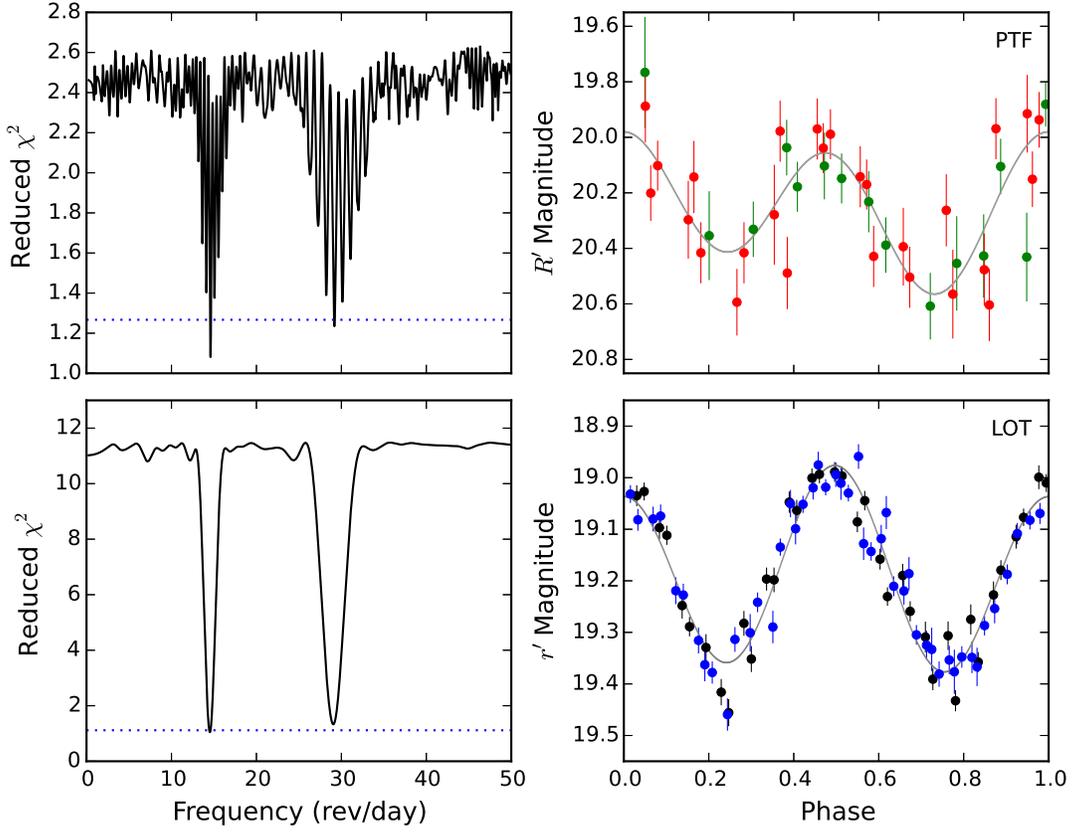}
\caption{Periodograms (left) and folded light curves (right) for 2005~EC$_{127}$ obtained from iPTF (upper) and LOT (lower) observations. The blue dotted lines in the periodograms indicate the uncertainties in the derived rotation periods. The grey lines in the light curves are the fitted results. The green and red filled circles in the PTF light curve are data points obtained from Feb 25 and 26, 2015, respectively. The black and blue filled circles in the LOT light curve are data points of the even and odd rotation cycles, respectively.}
\label{lc}
\end{figure}

The rotation period of $1.64 \pm 0.01$~hours (i.e., 14.6 rev/day) of 2005~EC$_{127}$ was first identified using the PTF light curve \citep{Chang2016}. Although the derived frequency of 14.6 rev/day is significant in the periodogram calculated from the PTF light curve, the corresponding folded light curve is relatively scattered (see upper panels of Fig.~\ref{lc}). Therefore, we triggered the follow-up observations using the LOT. The rotation periods of 2005~EC$_{127}$ derived from the LOT light curve is $1.65 \pm 0.01$~hours (i.e., 14.52 rev/day), which is in good agreement with the PTF result (see lower panels of Fig.~\ref{lc}). Both folded light curves show a clear double-peak/valley feature for asteroid  rotation (i.e., two periodic cycles). The peak-to-peak variations of the PTF and LOT light curves are $\sim 0.6$ and $\sim 0.5$~mag, respectively. This indicates that 2005~EC$_{127}$ is a moderately elongated asteroid and rules out the possibility of an octahedronal shape for 2005~EC$_{127}$, which would lead to a light curve with four peaks and an amplitude of $\Delta m < 0.4$~mag \citep{Harris2014}. Moreover, we cannot morphologically distinguish between the even and odd cycles in the LOT light curve. Therefore, we believe that 1.65 hours is the true rotation period for 2005~EC$_{127}$.

\section{Results and Discussion}
To estimate the diameter, $D$, of 2005~EC$_{127}$, we use:
\begin{equation}
  D = {1329 \over \sqrt{p_v}} 10^{-H/5}
\end{equation}
\citep[see][and references therein]{Harris2002}. Since the phase angle of the asteroid had a small change during our relatively short observation time span, the absolute magnitude of 2005~EC$_{127}$ is simply calculated using a fixed $G$ slope of 0.15 in the $H$--$G$ system \citep{Bowell1989}. We obtain $H_{R'} = 17.27 \pm 0.22$ and $H_{r'}17.30 \pm 0.02$~mag from the PTF and LOT observations, respectively\footnote{A $G$ slope of 0.24 for S-type asteroids \citep{Pravec2012} would make the $H$~magnitude $\sim 0.03$~mag fainter, which is equivalent to a $\sim 0.01$~km diameter difference, and within the uncertainty of our estimation.}. Because the absolute magnitude derived from the LOT observation has a smaller dispersion, we finally adopt $H_{r'} = 17.30$~mag for 2005~EC$_{127}$. We use $(V-R) = 0.516$ in the conversion of $H_{r'}$ to $H_V$ \citep{DeMeo2009, Pravec2012}, and then obtain $H_V = 17.82$ for 2005~EC$_{127}$. Assuming an albedo value of $p_v = 0.36 \pm 0.10$ for V-type and $p_v = 0.19 \pm 0.03$ for A-type asteroids \citep{Masiero2011, DeMeo2013}, diameters of $D \sim 0.6 \pm 0.1$ and $\sim 0.8 \pm 0.1$~km, respectively, are estimated for 2005~EC$_{127}$, where the uncertainty includes the residuals in light-curve fitting and the range of assumed albedos. Even when an extreme albedo value of $p_v = 1.0$ is applied, a diameter of 0.4 km is still obtained for 2005~EC$_{127}$. Since A-type asteroids are relatively uncommon in the inner main belt, we therefore assume an V-type asteroid for 2005~EC$_{127}$ in the following discussion. As shown in Fig.~\ref{dia_per}, 2005~EC$_{127}$ lies in the rubble-pile asteroid region and has a rotation period shorter than 2 hours. Therefore, we conclude that 2005~EC$_{127}$ is a large SFR.

\begin{figure}
\plotone{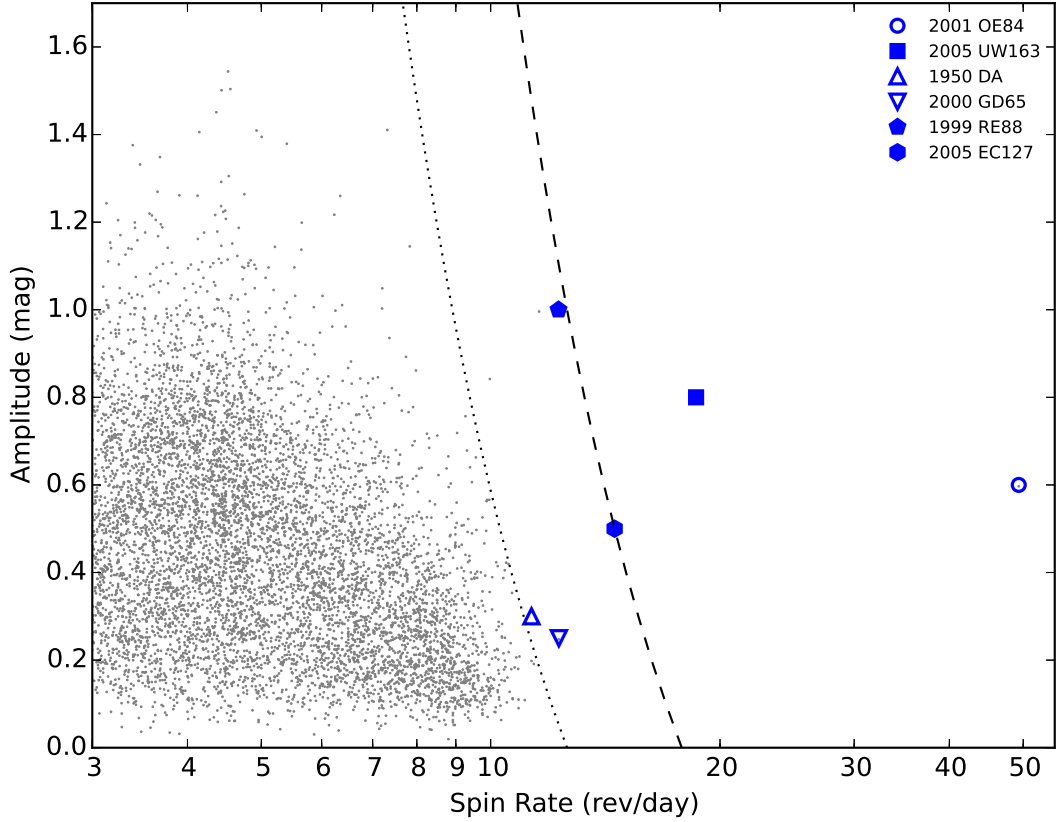}
\caption{Asteroid light-curve amplitude vs.\ spin rate. The gray dots are LCDB objects with reliable rotation periods. The blue symbols denote the six reported large SFRs, where the filled symbols were discovered by the iPTF. The dashed and dotted lines represent the estimated spin rate limits for rubble-pile asteroids of bulk densities of $\rho =$ 6 and 3~g~cm$^{-3}$, respectively, using $P \sim 3.3 \sqrt{(1+\Delta m)/\rho}$ hours, where  $\Delta m$ is the light-curve amplitude \citep{Harris1996}. Assuming a rubble-pile structure, the estimated bulk density of 2005~EC$_{127}$ (blue hexgon) would be $\sim 6$~g~cm$^{-3}$.}
\label{spin_amp}
\end{figure}

\begin{figure}
\plotone{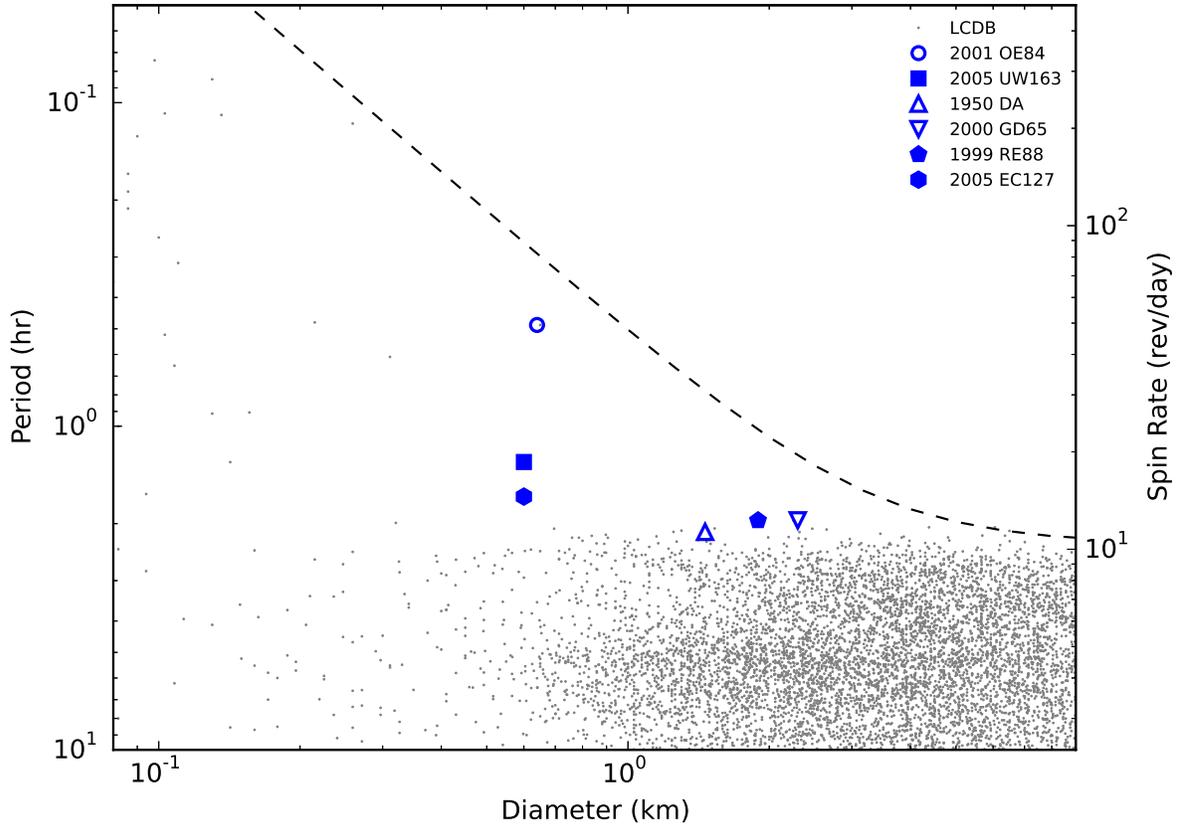}
\caption{Asteroid rotation period (spin rate) vs.\ diameter. The symbol assignments are the same as used in Fig.~\ref{spin_amp}. The large SFRs have somewhat smaller periods than the spin barrier at 2.2 hours. The dashed line is the predicted spin limit with cohesion $\kappa = k\bar{r}^{\,-1/2}$, where the strength coefficient $k = 2.25 \times 10^7$~dynes~cm$^{-3/2}$ and $\bar{r}$ is the mean radius \citep{Holsapple2007}.}
\label{dia_per}
\end{figure}

If 2005~EC$_{127}$ is a rubble-pile asteroid, a bulk density of $\rho \sim 6$~g~cm$^{-3}$ would be required to withstand its super-fast rotation (see Fig.~\ref{spin_amp}). This would suggest that 2005~EC$_{127}$ is a very compact object, i.e., comprised mostly of metal. However, such high bulk density is very unusual among asteroids. Moreover, 2005~EC$_{127}$ is probably an V-type asteroid. Therefore, this is a very unlikely scenario indeed.

Another possible explanation for the super-fast rotation of 2005~EC$_{127}$ is that it has substantial internal cohesion \citep{Holsapple2007, Sanchez2014}. Using the Drucker-Prager yield criterion\footnote{The detailed calculation is given in \citet{Chang2016}. This method has been widely used, e.g., in \citet{Holsapple2007, Rozitis2014, Polishook2016}.}, we can estimate the internal cohesion for asteroids. Assuming an average $\rho = 1.93$~g/cm$^3$ for V-type asteroids \citep{Carry2012}, a cohesion of $47 \pm 20$~Pa results for 2005~EC$_{127}$\footnote{For an A-type asteroid with average density $\rho = 3.73$~g/cm$^3$ \citep{Carry2012}, the cohesion would be 52~Pa.}. This modest value is comparable with that of the other large SFRs (see Table~\ref{known_sfr}), and also nearly in the cohesion range of lunar regolith, i.e., 100-1000~Pa~\citep{Mitchell1974}.

As shown by \citet{Holsapple2007}, the size-dependent cohesion would allow large SFRs to be present in the transition zone between monolithic and rubble-pile asteroids. However, only six large SFRs have been reported to date (including this work). This number is very small when compared with the number of large fast rotators (i.e., 738 objects in the LCDB). The reason for the rarity in detecting large SFRs from previous studies (i.e., the sparse number of large SFRs in the transition zone in Fig.~\ref{dia_per}) could be that: (a) The rotation periods are difficult to obtain for large SFRs due to their small diameters (i.e., faint brightness); or (b) The population size of large SFRs is intrinsically small. Therefore, a survey of asteroid rotation period with a larger sky coverage and deeper limiting magnitude can help to resolve the aforementioned question. If it is the latter case, these large SFRs might be monoliths, which have relatively large diameters and unusual collision histories.

We also note that none of the six reported large SFRs are classified as C-type asteroids. Therefore, any discovery of a large C-type SFR would fill out this taxonomic vacancy and help to understand the formation of large SFRs. In addition, the determination of the upper limit of SFR diameter is also important for understanding asteroid interior structure, since this can constrain the upper limit of internal cohesion of asteroids.

\section{Summary and Conclusions}
(144977) 2005~EC$_{127}$ is consistent with an V-/A-type inner-main-belt asteroid, based on our follow-up spectroscopic observations, with a diameter estimated to be $0.6 \pm 0.1$~km from the standard brightness/albedo relation. Its rotation period was first determined to be $1.64 \pm 0.01$~hours from our iPTF asteroid rotation-period survey, and then confirmed as $1.65 \pm 0.01$~hours by the follow-up observations reported here using the LOT. We categorize 2005~EC$_{127}$ as a large SFR, given its size and since its rotation period is less than the 2.2-hour spin-barrier.

Considering its 0.6 km diameter, 2005~EC$_{127}$ is most likely a rubble-pile asteroid. For 2005~EC$_{127}$ to survive under its super-fast rotation, either an internal cohesion of $47 \pm 20$~Pa or an unusually high bulk density of $\rho \sim 6$~g/cm$^3$ is required. However, the latter case is very unlikely for large asteroids, and more so for V-/A-type asteroids, as 2005~EC$_{127}$ has been classified. Only six large SFRs have been reported in the literature, including 2005~EC$_{127}$, the subject of this work. This number is very small compared with the number of existing large fast rotators. Therefore, future surveys will help to reveal whether this rarity in detection is due to the intrinsically small population size of large SFRs. Moreover, none of the known super-fast rotators have been classified as C-type asteroids, and the discovery of a large super-fast rotator of this type in future work would be an interesting development to further our understanding of the formation of large super-fast rotators.

\acknowledgments This work is supported in part by the Ministry of Science and Technology of Taiwan under grants MOST 104-2112-M-008-014-MY3, MOST 104-2119-M-008-024 and MOST 105-2112-M-008-002-MY3, and also by Macau Science and Technology Fund No. 017/2014/A1 of MSAR. We are thankful for the indispensable support provided by the staff of the Lulin Observatory and the staff of the Palomar Observatory. We thank the anonymous referee for his useful suggestions and comments.

\begin{deluxetable}{llcccccccccc}
\tabletypesize{\scriptsize} \tablecaption{Observational details.\label{obs_log}} \tablewidth{0pt}
\tablehead{\colhead{Telescope} & \colhead{Date} & \colhead{Filter} & \colhead{RA ($^\circ$)} & \colhead{Dec. ($^\circ$)} & \colhead{N$_\textrm{exp}$} & \colhead{$\Delta t$ (hours)} & \colhead{$\alpha$ ($^\circ$)} & \colhead{$r$ ($au$)} & \colhead{$\Delta$ ($au$)} & \colhead{$m$ (mag)} & \colhead{$H$ (mag)}} \startdata
  PTF & Feb 25--26 2015 & $R'$                         &  154.04 &  10.12 & 43 & 28.3 & 1.3 & 2.45 & 1.46 & 20.3 & 17.3\\
  LOT & Sept 24 2016   & $r'$                         &   23.81 &   2.81 & 84 &  7.3 & 2.5 & 2.03 & 1.03 & 19.2 & 17.3\\
 P200 & Oct 4 2016     & Spec.: $0.4-–0.9~\mu$m       &   23.65 &   2.21 &  3 &  0.5 & 7.7 & 2.05 & 1.07 & 19.5 &     \\
\enddata
\tablecomments{$\Delta t$ is observation time span and N$_\textrm{exp}$ is the total number of exposures.}
\end{deluxetable}

\begin{deluxetable}{rllcccccrcrrrl}
\tabletypesize{\scriptsize} \tablecaption{Confirmed large SFRs to date.\label{known_sfr}} \tablewidth{0pt}
\tablehead{\colhead{} & \colhead{Asteroid} & \colhead{Tax.} & \colhead{Per.} & \colhead{$\Delta m$} & \colhead{Dia.} & \colhead{$H$} & \colhead{Coh.} & \colhead{$a$} & \colhead{$e$} & \colhead{$i$} & \colhead{$\Omega$} & \colhead{$\omega$} & \colhead{Ref.}\\
	
\colhead{} & \colhead{} & \colhead{} & \colhead{(hours)} & \colhead{(mag)} & \colhead{(km)} & \colhead{(mag)} & \colhead{(Pa)} & \colhead{($au$)} & \colhead{} & \colhead{($^\circ$)} & \colhead{($^\circ$)} & \colhead{($^\circ$)} & \colhead{}	
	
	} \startdata
  (144977) & 2005~EC$_{127}$ & V/A & $1.65 \pm 0.01$ &  0.5     & $0.6 \pm 0.1$ & $17.8 \pm 0.1$ & $47 \pm 30$   & 2.21 &  0.17 &  4.75 & 336.9 & 312.8 & This work           \\
  (455213) & 2001~OE$_{84}$  & S & $0.49 \pm 0.00$ &  0.5     & $0.7 \pm 0.1$ & $18.3 \pm 0.2$ & $\sim1500^b$  & 2.28 &  0.47 &  9.34 &  32.2 &   2.8 & \citet{Pravec2002}   \\
  (335433) & 2005~UW$_{163}$ & V & $1.29 \pm 0.01$ &  0.8     & $0.6 \pm 0.3$ & $17.7 \pm 0.3$ & $\sim200^b$   & 2.39 &  0.15 &  1.62 & 224.6 & 183.6 & \citet{Chang2014b}   \\
   (29075) & 1950~DA         & M & $2.12 \pm 0.00$ &  0.2$^a$ & $1.3 \pm 0.1$ & $16.8 \pm 0.2$ & $64 \pm 20$   & 1.70 &  0.51 & 12.17 & 356.7 & 312.8 & \citet{Rozitis2014}  \\
   (60716) & 2000~GD$_{65}$  & S & $1.95 \pm 0.00$ &  0.3     & $2.0 \pm 0.6$ & $15.6 \pm 0.5$ & 150--450      & 2.42 &  0.10 &  3.17 &  42.1 & 162.4 & \citet{Polishook2016} \\
   (40511) & 1999~RE$_{88}$  & S & $1.96 \pm 0.01$ &  1.0     & $1.9 \pm 0.3$ & $16.4 \pm 0.3$ & $780 \pm 500$ & 2.38 &  0.17 &  2.04 & 341.6 & 279.8 & \citet{Chang2016}   \\
\enddata
\tablecomments{The orbital elements were obtained from MPC website, http://www.minorplanetcenter.net/iau/mpc.html.}
\tablenotetext{a}{$\Delta m$ is adopted from \citet{Busch2007}.}
\tablenotetext{b}{The cohesion is adopted from \citet{Chang2016}.}
\end{deluxetable}

\end{document}